\def\presentation{
\voffset -.50in  \hoffset -.19in
\oddsidemargin 0in \evensidemargin 0in
\marginparwidth .75in \marginparsep 7pt \topmargin 0in
\headheight 12pt \headsep .25in
\footheight 18pt \footskip .35in
\textheight 9.5in \textwidth 6.5in
\columnsep 10pt \columnseprule 0pt }
\documentstyle{article}
\presentation
\begin{document}
%

%
\def\tilde{\widetilde}
\def\bar{\overline}
\def\hat{\widehat}
\def\*{\star}
\def\[{\left[}
\def\]{\right]}
\def\({\left(}
\def\){\right)}
\def\zb{{\bar{z} }}
\def\frac#1#2{{#1 \over #2}}
\def\inv#1{{1 \over #1}}
\def\half{{1 \over 2}}
\def\d{\partial}
\def\der#1{{\partial \over \partial #1}}
\def\dd#1#2{{\partial #1 \over \partial #2}}
\def\vev#1{\langle #1 \rangle}
\def\bra#1{{\langle #1 |  }}
\def\ket#1{ | #1 \rangle}
\def\rvac{\hbox{$\vert 0\rangle$}}
\def\lvac{\hbox{$\langle 0 \vert $}}
\def\2pi{\hbox{$2\pi i$}}
\def\e#1{{\rm e}^{^{\textstyle #1}}}
\def\grad#1{\,\nabla\!_{{#1}}\,}
\def\dsl{\raise.15ex\hbox{/}\kern-.57em\partial}
\def\Dsl{\,\raise.15ex\hbox{/}\mkern-.13.5mu D}
\def\comm#1#2{ \BBL\ #1\ ,\ #2 \BBR }
\def\x{\stackrel{\otimes}{,}}
\def\det{ {\rm det}}
\def\tr{{\rm tr}}
%
%
\def\th{\theta}		\def\Th{\Theta}
\def\ga{\gamma}		\def\Ga{\Gamma}
\def\be{\beta}
\def\al{\alpha}
\def\ep{\epsilon}
\def\la{\lambda}	\def\La{\Lambda}
\def\de{\delta}		\def\De{\Delta}
\def\om{\omega}		\def\Om{\Omega}
\def\sig{\sigma}	\def\Sig{\Sigma}
\def\vphi{\varphi}
%
%
\def\CA{{\cal A}}	\def\CB{{\cal B}}	\def\CC{{\cal C}}
\def\CD{{\cal D}}	\def\CE{{\cal E}}	\def\CF{{\cal F}}
\def\CG{{\cal G}}	\def\CH{{\cal H}}	\def\CI{{\cal J}}
\def\CJ{{\cal J}}	\def\CK{{\cal K}}	\def\CL{{\cal L}}
\def\CM{{\cal M}}	\def\CN{{\cal N}}	\def\CO{{\cal O}}
\def\CP{{\cal P}}	\def\CQ{{\cal Q}}	\def\CR{{\cal R}}
\def\CS{{\cal S}}	\def\CT{{\cal T}}	\def\CU{{\cal U}}
\def\CV{{\cal V}}	\def\CW{{\cal W}}	\def\CX{{\cal X}}
\def\CY{{\cal Y}}	\def\CZ{{\cal Z}}
%
%
\font\numbers=cmss12
\font\upright=cmu10 scaled\magstep1
\def\stroke{\vrule height8pt width0.4pt depth-0.1pt}
\def\topfleck{\vrule height8pt width0.5pt depth-5.9pt}
\def\botfleck{\vrule height2pt width0.5pt depth0.1pt}
\def\Zmath{\vcenter{\hbox{\numbers\rlap{\rlap{Z}\kern
		0.8pt\topfleck}\kern
		2.2pt \rlap Z\kern 6pt\botfleck\kern 1pt}}}
\def\Qmath{\vcenter{\hbox{\upright\rlap{\rlap{Q}\kern
                   3.8pt\stroke}\phantom{Q}}}}
\def\Nmath{\vcenter{\hbox{\upright\rlap{I}\kern 1.7pt N}}}
\def\Cmath{\vcenter{\hbox{\upright\rlap{\rlap{C}\kern
                   3.8pt\stroke}\phantom{C}}}}
\def\Rmath{\vcenter{\hbox{\upright\rlap{I}\kern 1.7pt R}}}
\def\Z{\ifmmode\Zmath\else$\Zmath$\fi}
\def\Q{\ifmmode\Qmath\else$\Qmath$\fi}
\def\N{\ifmmode\Nmath\else$\Nmath$\fi}
\def\C{\ifmmode\Cmath\else$\Cmath$\fi}
\def\R{\ifmmode\Rmath\else$\Rmath$\fi}
\def\cadremath#1{\vbox{\hrule\hbox{\vrule\kern8pt\vbox{\kern8pt
			\hbox{$\displaystyle #1$}\kern8pt} 
			\kern8pt\vrule}\hrule}}
\def\proof{\noindent {\underline {Proof}.} }
\def\cqfd{ {\hfill{$\Box$}} }
\def\square{ {\hfill \vrule height6pt width6pt depth1pt} } 
%
%
\def\debut{ \begin{eqnarray} }
\def\fin{ \end{eqnarray} }
\def\non{ \nonumber }
%

%
%
\rightline{SPhT-99-012}
  ~\vskip 1cm
\centerline{\LARGE On The Three Point Velocity Correlation Functions}
\bigskip
\centerline{\LARGE in 2d Forced Turbulence.}
\vskip 1cm
\vskip1cm
\centerline{\large  Denis Bernard
\footnote[1]{Member of the CNRS; dbernard@spht.saclay.cea.fr} }
\centerline{Service de Physique Th\'eorique de Saclay
\footnote[2]{\it Laboratoire de la Direction des Sciences de la
Mati\`ere du Commisariat \`a l'Energie Atomique.}}
\centerline{F-91191, Gif-sur-Yvette, France.}
\vskip2cm
Abstract.\\
We present a simple exact formula for the three point velocity correlation
functions in two dimensional turbulence which is valid on all scales
and which interpolates between the direct and inverse cascade regimes.
As expected, these correlation functions are universal in these extreme regimes.
We also discuss the effect of anisotropy and friction.
%

%
%
\vfill
\newpage
%

%
%
%
The aim of this short note is to write down an explicit formula for the three point 
velocity correlation functions in 2d turbulence. See eqs.(\ref{3point}) below and its
consequences. This formula differs from the usual Kolmogorov formula,
cf eg ref.\cite{frisch}, by the fact that it incorporates the existence of two 
inertial ranges which respectively correspond to the inverse 
energy cascade and the direct enstrophy cascade.
Although expected on scaling ground, this formula, and its simple proof,  
was surprisingly never spelled out in the turbulent literature.
It is nevertheless one of the rare exact results on 2d turbulence.
We thus feel that it was worth making it more public.
\bigskip

{\bf Kraichnan's scaling theory.}
A special feature which distinguishes two dimensional
from three dimensional fluid mechanics is 
the conservation of vorticity moments in the inviscid limit. 
As first pointed out by Kraichnan in a remarkable paper \cite{kraich},
this opens the possibility for quite different scenario 
for the behavior of turbulent flows in two and three dimensions.
In two dimensions the inviscid Navier-Stokes equation admits
two quadratic conserved quantities: the energy $\int \frac{u^2}{2}$
and the enstrophy $\int \frac{\om^2}{2}$ with $u$ the velocity
and $\om$ the vorticity. As argued by Kraichnan, if energy and enstrophy
density are injected at a scale $L_i$, with respective rate $\bar \ep$ and 
$\bar \ep_w\simeq \bar \ep L_i^{-2}$, the turbulent system should
react such that the energy flows toward the large scales and the enstrophy
towards the small scales. As this energy flow is quite the opposite
to the one involved in Kolmogorov's picture for 3d turbulence, 
one usual refers to the infrared energy flow as the inverse 
cascade and to the ultraviolet enstrophy flow as the direct cascade. 
The fact that energy has to escape to the large scales may be
understood from the fact that in absence of forcing the time variation
of the energy is $\d_t \int \frac{u^2}{2} = -\nu \int \frac{\om^2}{2}$
with $\nu$ the viscocity. It thus vanishes in the inviscid limit
$\nu\to 0$ if the enstrophy remains finite and the energy cannot
be dissipated at small scales.

In the (IR) inverse cascade, scaling arguments lead to Kolmogorov's spectrum,
with $E(k)\sim \bar \ep^{2/3}\,k^{-5/3}$ for the energy and 
$(\De u)(r) \sim (\bar \ep r)^{1/3}$
for the variation of the velocity on scale $r$.

In the (UV) direct cascade, scaling arguments give Kraichnan's spectrum
with $E(k)\sim \bar \ep_w^{2/3}\, k^{-3}$ for the energy and 
$(\De u)(r) \sim (\bar \ep_w r^3)^{1/3}$ for the velocity variation.

Of course the direct and inverse cascade have been extensively analysed,
both numerically, see eg. \cite{numer} and refs. therein
for an (incomplete) sample of references,
and theoretically, see eg. \cite{theo,eyink} 
and refs therein for a few relevent references,
some of which discussing logarithmic corrections to Kraichnan's scaling.
More recently, the inverse cascade has been observed
experimentally, as described in ref.\cite{tabel}. Within experimental precision
it shows no deviation from Kolmogorov's scaling.

\bigskip
  
{\bf A model and its hypothesis.}
As usual, to statistically model turbulent flows we consider 
the Navier-Stokes equation with an extra forcing term.
Let $u^j(x,t)$ be the velocity field for an incompressible fluid, $\nabla\cdot u=0$.
In two dimensions the incompressibility implies that $u(x,t)$ derives from a stream 
function $\Phi$ such that $u^k=\ep_{kj}\d_j\Phi$ with $\ep_{kj}$
the antisymmetric tensor.  The Navier-Stockes equation reads:
\debut
\d_t u^j + (u\cdot\nabla) u^j - \nu \nabla^2 u^j = -\nabla^jp + f^j
\label{nseq}
\fin
with $p$ the pressure and $f(x,t)$ the external force such 
that $\nabla\cdot f=0$. We choose the force to
be Gaussian, white-noice in time, with zero mean and two point function:
\debut
\vev{f^j(x,t)\ f^k(y,s)} = C^{jk}(x-y)\ \de(t-s) \label{vvf}
\fin 
where $C^{jk}(x)$, with $\nabla^j\, C^{jk}(x)=0$, 
is a smooth function varying on a scale $L_i$,
fastly decreasing at infinity and regular at the origin.
The scale $L_i$ represents the injection length.
We shall assume translation, rotation and parity invariance, unless
otherwise specified.  Let  $\hat C(x)\equiv trC(x)$. 
Its Taylor expansion at the origin will be denoted as: 
$\hat C(x)= 2\bar \ep - \bar \ep_w\ r^2/2 + \cdots$ with $r^2=x^kx_k$.
The transversality condition $\nabla^j\, C^{jk}(x)=0$ ensures
that $\hat C = \nabla^k \Th^k$  with
$\Th^k(x)= \bar \ep x^k - \bar \ep_w\ x^k r^2/8 + \cdots$ at short distances.
Physical interpretation of $\bar \ep$ and $\bar \ep_w$ 
will be given later.

The vorticity $\om$, with $\om=\ep_{ij}\d_iu_j$, is transported by the fluid:
\debut
\d_t\om +(u\cdot\nabla)\om - \nu \nabla^2 \om = F \label{omeq}
\fin
with $F=\ep_{ij}\d_if_j$. The correlation function of the 
vorticity forcing term is thus:
\debut
\vev{F(x,t)\ F(y,s)} = G(x-y)\ \de(t-s) \label{vvF}
\fin
with  $G=-\nabla^2 \hat C$. In particular $G(0)= 2\bar \ep_w$.
The fact that the correlation function of
the vorticity forcing is a gradiant will have physical consequences.
Physically eq.(\ref{omeq}) means that for smooth solutions
any powers of the vorticity, and in particular the enstrophy 
$\int \frac{\om^2}{2}$, is conserved in absence of viscocity and forcing. 

\medskip
Since the inviscid limit is of course not under analytical control,
we have to make a few hypothesis which encode Kraichnan's scenario
of inverse and direct cascades.
The hypothesis are the following:\\
(i) the velocity correlation functions are assumed to be smooth at finite
viscocity and correlations of the velocity (without derivatives
but at points coinciding or not) exist in the inviscid limit;\\
(ii) Galilean invariant correlation functions, and in particular 
the velocity structure functions which are correlations
of differences of the velocity field, are stationary;\\
(iii) last but not least, in agreement with Kraichnan's picture,
we demand that energy dissipative anomalies (but not
enstrophy dissipative anomalies) be absent.

The two first hypothesis are standard in statistical approach to
turbulence while the third one is special to two dimensions.
It follows by demanding that the enstrophy density $\Om=\frac{\om^2}{2}$ is finite
in the inviscid limit since the mean enstrophy density times the viscosity
is equal to the mean dissipation rate, 
$\nu \vev{\Om} = \frac{\nu}{2} \vev{(\nabla u)\cdot (\nabla u)}$.

\bigskip

{\bf Velocity correlations.}
Let us look at the two point velocity correlation function $\vev{u(x)\cdot u(0) }$.
As is well known, it satisfies the following equation at finite viscocity:
\debut
\d_t\vev{u(x)\cdot u(0) } 
-\half \nabla_x^k \vev{ (\De u^k)(x)\ (\De u)^2(x)} 
+2\nu \vev{(\nabla u)(x)\cdot (\nabla u)(0) } = \hat C(x)
\label{hopf}
\fin
Here and in the following we shall denote velocity
differences by $(\De u^k)(x)\equiv u^k(x)-u^k(0)$.
Eq.(\ref{hopf}) assumes translation invariance and uses the fact 
that the external force is Gaussian and white-noice in time.
Thanks to the fluid incompressibility  
the pressure drops out from this equation.
The strategy consists in taking various limits of eq.(\ref{hopf}) 
in various orders.  Let us take first the limit $x\to 0$ 
followed by the inviscid limit. In this limit the second term in eq.(\ref{hopf})
vanishes due to the assumed smoothness of the correlation functions,
hypothesis (i).  Recall now the hypothesis (iii) concerning the absence of
energy dissipation. It in particular means that:
\debut
\lim_{\nu\to 0} \lim_{x\to 0} 
\nu \vev{ (\nabla u)(x)\cdot (\nabla u)(0) } = 0 
\fin
Therefore the third term in eq.(\ref{hopf}) also vanishes. 
It implies that in the inviscid limit the mean energy increases with time according to:
\debut
\d_t \vev{\frac{u^2}{2}}_{\nu=0} = \half \hat C(0) =  \bar \ep 
\label{energy}
\fin
Thus $\vev{\frac{u^2}{2}}_{\nu=0} =  \bar \ep\, t$ up to a constant,
and $\bar \ep$ is indeed the energy injection rate.
This is simply the obvious statement
that in absence of energy dissipation, and/or in absence of
friction or other processes by which the energy may escape,
all energy injected into the system is transfered to the fluid.
It is expected to be transfered to the mode with the smallest possible
momentum, the so-called condensate \cite{kraich}. 
In particular eq.(\ref{energy}) shows that in absence of energy dissipation
a stationary state can not be reached although structure
functions may converge at large time.
This is one important difference between 2d and 3d turbulence.
\medskip

Let us now assume that the two point structure function 
is stationary, ie. $\d_t\vev{ (\De u)^2(x)}=0$, hypothesis (ii).
  From eq.(\ref{hopf}) one obtains in the inviscid limit:
\debut
\half\nabla_x^k \vev{ (\De u^k)(x)\ (\De u)^2(x)}_{\nu=0} = 2\bar \ep - \hat C(x)
\label{3point0}
\fin
Integrating it using parity invariance gives:
\debut
\vev{ (\De u^k)(x)\ (\De u)^2(x)}_{\nu=0} = 2\({\bar \ep x^k - \Th^k(x) }\)
\label{3point}
\fin
with $\nabla_x\cdot \Th(x)= \hat C(x)$.
Eq.(\ref{3point}) together with eq.(\ref{exactom}) below fully determine
the three point velocity correlation.
Although simple to derive, this equation seems not to have appeared in the
existing literature \footnote{ We thank G. Falkovich and R. Kraichnan for
informations on the literature on turbulence.}.

Eq.(\ref{3point}) in particular shows that the inverse energy cascade
takes place only if there is no dissipation anomaly,
and thus only if the non-galilean invariant velocity correlation 
functions do not reach a stationary state (in absence of friction). 
Of course this is also a direct consequence of the physical fact that the
energy condensates into the mode of smallest possible momemtum.
As expected, eq.(\ref{3point})  yields to Kolmogorov's scaling
at large scale, since there $\Th^k(x)$ vanishes, and Kraichnan's
scaling at small scale since $\({\bar \ep x^k - \Th^k(x) }\)\sim r^3$ 
at short distance.  But one could be a little more precise.
\medskip

Let us first consider the short distance behavior in which
the Kraichnan's direct cascade takes place. 
This corresponds to scale much smaller than the injection length $x<<L_i$.
There, $$\vev{ (\De u^k)(x)\ (\De u)^2(x)}\simeq \inv{4}\bar \ep_w x^k x^2$$
Assuming isotropy and parity invariance the
three point functions will be linear combinations
of terms proportional to $x^ix^jx^k$ or 
$(\de^{ij}x^k+\de^{jk}x^i+\de^{ki}x^j)x^2$. 
Among these two proportionality coefficients only one of them 
could be fixed using eq.(\ref{3point}) only. 
However,  the other coefficient is
fixed by using another exact result for correlation
functions mixing the vorticity and the velocity which is described in
the following, see eq.(\ref{exactom}).  This then gives  for $x\to 0$:
\debut
\vev{(\De u^i)(x) (\De u^j)(x) (\De u^k)(x) }
\simeq \frac{\bar \ep_w}{8}\({ 
(\de^{ij}x^k+\de^{jk}x^i+\de^{ki}x^j)x^2 - 2 x^ix^jx^k }\)
\label{3pointbis}
\fin
For the transverse and longitudinal correlations this becomes:
\debut
\vev{ (\De u)_{\|}^3 } = \vev{ (\De u)_{\|} (\De u)_{\bot}^2 }
\simeq +\frac{{\bar \ep_w}}{8}\ r^3
\label{law}
\fin
The coefficient $\bar \ep_w$ will be shown to be equal to the
mean enstrophy dissipation rate in the following.
Thus, as expected the 3-point velocity functions, which only depend
on the enstrophy injection rate, are universal in the direct
cascade. Eq.(\ref{law}) may be called the "$+1/8$ law" of the direct cascade.
\medskip

Consider now the large distance behavior in which the
Kraichnan's inverse cascade takes place, ie. scale $x>>L_i$. There,
$$\vev{ (\De u^k)(x)\ (\De u)^2(x)}\simeq 2\bar \ep x^k$$
There is two possible terms for the three point 
functions, $(\de^{ij}x^k+\de^{jk}x^i+\de^{ki}x^j)$ and
$(x^ix^jx^k)/x^2$ whose coefficients can not be completely fixed
using only eq.(\ref{3point}). But again these will be fixed
by looking at the vorticity correlation functions, see below eq.(\ref{exactom}).
One then gets for $x\to \infty$:
\debut
\vev{(\De u^i)(x) (\De u^j)(x) (\De u^k)(x) }
\simeq \frac{\bar \ep}{2} (\de^{ij}x^k+\de^{jk}x^i+\de^{ki}x^j)
\label{kol}
\fin
with $\bar \ep$ the mean energy injection rate.
Of course this gives the "$+3/2$ law" for the longitudinal
statistics in the inverse cascade:
\debut
\vev{ (\De u)_{\|}^3 } =3\vev{ (\De u)_{\|} (\De u)_{\bot}^2 }
\simeq +\frac{3\bar \ep}{2} r 
\label{lawbis}
\fin
This law and Kolmorgorov's scaling as well as the existence of a condensate
in which the energy is accumulating was experimentally verified in \cite{tabel}.
\bigskip

{\bf Vorticity correlations.}
We now establish a formula for a mixed correlation function involving 
the vorticity and the velocity.
Assuming that the structure functions of the velocity 
reach a stationary state implies that correlations
of the vorticity also become stationary. 
The stationarity condition for the two point vorticity
functions, ie. $\d_t \vev{\om(x)\om(y)}=0$, implies:
\debut
-\half \nabla_x^k \vev{ (\De u^k)(x)\ (\De \om)^2(x)}
+2 \nu \vev{ \nabla \om(x)\cdot \nabla\om(0) } = G(x)
\label{om1}
\fin
with $(\De\om)(x)=\om(x)-\om(0)$. As for the velocity correlations,
let us first take the  limit of coincident point $x\to 0$
at finite viscocity. The first term in eq.(\ref{om1})
then vanishes by the hypothesis on the smoothness of the correlation
functions at finite $\nu$. Taking then the inviscid limit
leads to:
\debut
\lim_{\nu \to 0} \nu \vev{ (\nabla \om)^2(x) }
= \half G(0) = \bar \ep_w 
\label{omdis}
\fin
This is just the usual statement on enstrophy dissipative
anomaly. It equals the enstrophy dissipation rate and the enstrophy
injection rate.  Let us now take the limits in the reversed order,
the inviscid limit first.  In that case the second term of eq.(\ref{om1}) 
drops out and one gets:
\debut
-\half \nabla_x^k \vev{ (\De u^k)(x)\ (\De \om)^2(x)}_{\nu=0}
=G(x)  \non
\fin
Assuming isotropy and parity invariance this gives:
\debut
\vev{ (\De u^k)(x)\ (\De \om)^2(x)}_{\nu=0}
= 2 \frac{x^k}{r} (\frac{d\hat C}{dr})(r) = - 2\bar \ep_w x^k + O(x^2)
\label{exactom}
\fin
Thus this correlation is universal in the (UV) direct cascade regime.
Its behavior at large scale depends on the way the forcing decreases at infinity.
However the fact that these correlations decrease faster than $O(1/r)$
at infinity is linked to the fact the vorticity forcing correlation is a gradiant.
The ultraviolet behavior (\ref{exactom}) was also described in \cite{eyink}.
As mentioned above, this equation may be used to fix the coefficients 
of the infrared and ultraviolet expansions of the three point velocity
function left undertermined by eq.(\ref{3point}). (One should use
the relation $3\vev{(\De u^z)(\De \om)^2}=4\d_z^2\vev{(\De u^z)^3}$
where $u^z$ is the velocity component in complex coordinates $z=x+iy$.) 
But eq.(\ref{exactom}) alone would not have been enough to determine
these asymptotic expansions. 

\bigskip

{\bf Influence of anisotropy and parity symmetry breaking.}
Anisotropy is irrelevent both in the ultraviolet and in the infrared.
Indeed, let us model anisotropy by incorporating higher spin components 
in the forcing correlation function $C^{ij}(x)$, assuming that they
are still regular at the origin and decrease at infinity. 
These components will be subdominant in eq.(\ref{3point0})
both in the ultraviolet (since the spin $n$ component will
behave as $r^{n}$) and in the infrared (since they
also vanish at infinity).

Suppose now that parity symmetry may be broken. 
Eq.(\ref{3point}) is still valid (since it only assumes
translation invariance) except that $\Th^k(x)$ is determined
up to $\Th^k \to \Th^k + \ep^{kj} \d_j \Xi(r)$.
This may change the ultraviolet scaling of the transverse
velocity correlations but not the scaling of the longitudinal 
correlations $\vev{ (\De u)_{\|}^3 }$ although the amplitude
may be modified.

\bigskip

{\bf Influence of friction.}
In physical systems the infrared energy cascade will terminate 
at the largest possible scale at which the energy will escape.
This could be mimic by introducing a friction term in the
Navier-Stokes equation which then becomes:
\debut
\d_t u^j + (u\cdot\nabla) u^j - \nu \nabla^2 u^j + \inv{\tau}\, u^j= -\nabla^jp + f^j
\label{frict}
\fin
with $\tau$ the friction relaxation time, $\tau>0$.
Friction brings another inviscid characteristic length $L_f$ into the problem:
$L_f\simeq \tau^{3/2}\, \bar \ep^{1/2}$. It increases as the friction is reduced
and one may suppose $L_i<< L_f$. It is the length at which the energy is extracted.
The friction term dominates over the advection term at scale bigger than $L_f$.
So the direct cascade should take place at distances $x<<L_i<<L_f$ and the
inverse cascade at distances $L_i<<x<< L_f$.

Under the same hypothesis as before, the mean energy density
relaxes in the inviscid limit  according to
$\d_t \vev{\frac{u^2}{2} } + \inv{\tau} \vev{u^2 } = \bar \ep $.
It therefore reaches a stationary limit with $\vev{u^2 }=\bar \ep\,\tau$.
Stationarity of the two point structure function, ie. $\d_t \vev{(\De u)^2(x)}=0$,
then gives in the inviscid limit:
\debut
\half\nabla_x^k \vev{ (\De u^k)(x)\ (\De u)^2(x)}
+\inv{\tau} \vev{(\De u)^2(x)} = 2\bar \ep - \hat C(x)
\label{dissip}
\fin
Similarly, the stationarity of the vorticity correlations gives
an equation similar to eq.(\ref{om1}) but with an extra term
representing the friction. 
As for the case without friction let us first take the limit $x\to 0$
at finite viscocity and then the inviscid limit. Let us denote by
$\bar \ep_w = \half G(0)$ the enstrophy injection rate and by
$\hat \ep_w = \lim_{\nu\to 0} \nu\vev{(\nabla \om)^2(x)}$
the enstrophy dissipation rate. One then gets:
$\vev{\om^2} = \tau \({ \bar \ep_w - \hat \ep_w }\)$  with
$\vev{\om^2} = \lim_{\nu \to 0} \vev{\om^2(x)}$.
It simply means that the enstrophy density is equal to the
difference of the enstrophy injection and the enstrophy dissipation
rates times the friction relaxation time. In particular, if 
$\tau$ is finite so is the enstrophy density.
As a consequence the vorticity two point correlation function
$\vev{\om(x)\om(0)}$ will stay finite since it is bounded by $\vev{\om^2}$.
Taking the limit in the reversed order, first $\nu \to 0$,
yields to the inviscid stationary equation:
\begin{eqnarray}
-\half\nabla_x^k \vev{ (\De u^k)(x)\ (\De \om)^2(x)}
+\frac{2}{\tau} \vev{\om(x)\om(0)} = G(x)
\label{disom}
\end{eqnarray}

Let us look at small distances in 
which the friction should be irrelevent. 
Let $\bar \Om =\lim_{x\to 0}\vev{\om(x)\om(0)} $, which is expected
to be equal to $\tau(\bar \ep_w -\hat \ep_w)$ although nothing
prevent it to be different. Since $\bar \Om < \tau \bar \ep_w$, the
second term in the l.h.s. of eq.(\ref{disom}) cannot dominate
and $\nabla_x^k \vev{ (\De u^k)(x)\ (\De \om)^2(x)}\simeq {\rm const}.$
as $x\to 0$. This implies that the velocity three point function
$\vev{(\De u)^3}$ scale as $r^3$. 
In other words, Kraichnan's scaling of the three point function
is robust to friction in the direct cascade 
although the amplitude may change. 

More precisely, suppose that $\bar \Om$ is finite and non-vanishing. 
Then the scaling formula (\ref{3pointbis}) for the three
point function still holds but
with $\bar \ep_w$ replaced by $(\bar \ep_w\tau - \bar \Om)/\tau$.
Recall that it is likely that $(\bar \ep_w\tau - \bar \Om)/\tau$
is equal to the enstrophy dissipation rate $\hat \ep_w$,
meaning that in presence of friction one has
simply to replace the injection rate by the dissipation rate
in formula (\ref{3pointbis}). 
Moreover, the finiteness of the vorticity two point function
at coincident points also implies that:
$\vev{(\De u)^2(x)} \simeq \frac{\bar \Om}{2}\, r^2$ since
$\nabla^2_x \vev{(\De u)^2(x)}=2 \vev{\om(x)\om(0)}$.
It means that the UV energy spectrum is
\debut
E(k) \propto \bar \Om\, k^{-3}   \label{3spec}
\fin
without any logarithmic correction.
This scaling may be broken only if 
$\bar \Om=\lim_{x\to 0}\vev{\om(x)\om(0)}_{\nu=0}$ vanishes.
It is worth specifying more precisely in which scale domain eq.(\ref{3spec})
will be valid. At finite viscocity, there are two ultraviolet characteristic
lengths, the usual dissipative length $r_d \simeq \nu^{3/4}\, \ep^{1/4}$
and another friction length $l_f \simeq \nu^{1/2}\, \tau^{1/2}$
above which friction dominates over dissipation. Since $r_d<<l_f$ in the
limit we are considering, $\nu\to 0$ and $\tau$ fixed, eq.(\ref{3spec})
will be valid for $l_f << x << L_i$.

\medskip

Let us now consider distances larger than the injection length $L_i$ 
but smaller than the friction length $L_f$. Then,
$2\bar \ep - \hat C(x)\simeq 2\bar \ep$ and positivity
argument cannot be applied. However, unless miraculous cancelations
between the two terms in the l.h.s. of eq.(\ref{dissip}), (which would mean
that the domains in which advection or friction dominate intertwine), 
the correlation $\vev{ (\De u^k)(x)\ (\De u)^2(x)}$ will still scale as $r$.
But clearly this argument is less robust than the one used for
the short distance analysis. 

\bigskip

{\bf Conclusions.}
Besides giving the expected formula for the 3-point velocity correlation functions,
this short proof also indicates that if the inverse cascade takes place, as
experimentally verified, then, in absence of friction, the non-Galilean
invariant velocity correlation functions do not become stationary,
although structure functions do. However, 
it gives no hints on how to decipher the behavior of the vorticity, 
one of the main challanging problems of two dimensional turbulence.
\bigskip

{\bf Acknowledgements:} 
We thank G. Falkovich and R. Kraichnan for informative correspondences.


\begin{thebibliography}{}
%

\bibitem{frisch} U. Frisch, {\it  Turbulence: the legacy of A. Kolmogorov},
Cambridge University Press 1995.

\bibitem{kraich} R. Kraichnan, Phys. Fluids 10 (1967) 1417;\\
R. Kraichnan, J. Fluid. Mech 47 (1971) 525.

\bibitem{numer} See eg. and refs. therein: 
E. Siggia and H. Aref, Phys. Fluids 24 (1981) 171;\\
U. Frisch and P. Sulem Phys. Fluids 27 (1984) 1921;\\
A. Babiano, C. Basdevant, B. Legras, R. Sadourny, J. Fluid. Mech 183 (1987) 379;\\
B. Legras, P. Santangelo and R. Benzi, Europhys. Lett. 5 (1988) 37;\\
L. Smith and V. Yakhot, Phys. Rev. Let. 71 (1993) 352;\\
V. Borue, Phys. Rev. Let. 71 (1993) 3967;\\
A. Babiano, B. Dubrulle and P. Frick, Phys. Rev E52 (1995) 3719.

\bibitem{theo} See eg. the  following reviews:\\
R. Kraichnan and D. Montgomery, Rep. Prog. Phys. 43 (1980) 548;\\
H. Rose and P. Sulem, J. de Phys. 39 (1978) 441,\\
and more recently, see eg. and refs. therein:\\
M. Melander, N. Zabusky, J. Mc Williams, J. Fluid. Mech. 195 (1988) 303;\\
R. Benzi, G. Paladin, A. Vulpiani, Phys. Rev. A 42 (1990) 3654;\\
A. Polyakov, Nucl. Phys. B 396 (1993) 367;\\
G. Falkovich and V. Lebedev, Phys. Rev. E 50 (1994) 3885.

\bibitem{eyink} G. Eyink Physica D 91 (1996) 97.

\bibitem{tabel} J. Paret and P. Tabeling, Phys. Fluids, 10 (1998) 3126.


%
%
\end{thebibliography}
\end{document}